\documentclass[times,authoryear]{elsarticle}

\usepackage{jasr}
\usepackage{framed,multirow}

\usepackage{amssymb}
\usepackage{latexsym}

\usepackage{graphicx}
\usepackage{amsmath}
\usepackage{multirow}

\usepackage[switch]{lineno}

\usepackage{url}
\usepackage{xcolor}
\definecolor{newcolor}{rgb}{.8,.349,.1}
\newcommand{\highlight}[1]{#1}

\usepackage[citebordercolor=white]{hyperref}

\journal{Advances in Space Research}

\begin{document}

\verso{Affan Adly Nazri \textit{etal}}

\begin{frontmatter}

\title{Temporal variability in low-frequency radio interference: Insight from high-cadence monitoring at a candidate radio notification zone in Malaysia}

\author[1]{Affan Adly \snm{Nazri}}
\ead{affanadly.astro@gmail.com}
\author[1]{Zamri \snm{Zainal Abidin}\corref{cor1}}
\cortext[cor1]{Corresponding author: 
  Tel.: +0-000-000-0000;  
  fax: +0-000-000-0000;}
\ead{zzaa@um.edu.my}
\author[1]{Mohamad Ridhauddin \snm{Mat Sabri}}
\author[1,2]{Zulfazli \snm{Rosli}}
\author[1]{Mohd Shaiful Rizal \snm{Hassan}}
\author[1]{Mohd Shazwan \snm{Mohd Radzi}}
\author[1]{Ahmad Najwan \snm{Zulkiplee}}
\author[1]{Dalilah Nur Fathiah \snm{Hanim Razak}}
\author[1]{Norsyazwani \snm{Asmi}}
\author[3]{Jinsong \snm{Ping}}
\author[3]{Mingyuan \snm{Wang}}
\author[4,5]{Liang \snm{Dong}}

\affiliation[1]{organization={Center for Astronomy and Astrophysics Research},
                addressline={Department of Physics, Faculty of Science, Universiti Malaya},
                city={Kuala Lumpur},
                postcode={50603},
                country={Malaysia}}

\affiliation[2]{organization={Centre of Foundation, Language and Malaysian Studies},
                addressline={Universiti Malaya-Wales},
                city={Kuala Lumpur},
                postcode={50480},
                country={Malaysia}}
                
\affiliation[3]{organization={National Astronomical Observatories},
                addressline={Chinese Academy of Sciences},
                city={Beijing},
                postcode={100012},
                country={People's Republic of China}}

\affiliation[4]{organization={Yunnan Astronomical Observatory},
                addressline={Chinese Academy of Sciences},
                city={Kunming},
                postcode={650216},
                country={People's Republic of China}}
\affiliation[5]{organization={Yunnan Sino-Malaysian International Joint Laboratory of HF-VHF Advanced Radio Astronomy Technology},
                city={Kunming},
                postcode={650216},
                country={People's Republic of China}}

\received{xx xxx 2025}
\finalform{xx xxx 2025}
\accepted{xx xxx 2025}
\availableonline{xx xxx 2025}
\communicated{}

\begin{abstract}
Extensive radio frequency interference (RFI) monitoring is essential in the site selection process before constructing radio astronomy observatories, followed by mitigation strategies to minimize its adverse effects. Malaysia has an enormous prospect for radio astronomy due to its prominent location in the centre of Southeast Asia, but is challenged by its relatively high population density. In this research article, we perform high-cadence, low-frequency RFI monitoring at two sites, each representing an urban and a rural environment. Using modified generalized spectral kurtosis (GSK) as an RFI detection method, we ascertain the suitability of Glami Lemi, a rural area in the centre of Peninsular Malaysia previously assigned as a candidate radio notification zone (RNZ), as a potential site for radio astronomy observations due to its lower RFI contamination in our high-cadence monitoring, especially when compared with urban areas. We identified a number of persistent and transient RFI in our dataset, associate each of them with their potential origins and, if present, characterize their temporal evolution. A few types of RFI mitigation strategies were also tested and discussed. This study lays the groundwork for Malaysia’s endeavours in establishing its first research-grade radio telescope, emphasizing the importance of robust RFI detection and mitigation strategies in optimizing observational outcomes.
\end{abstract}


\end{frontmatter}


\section{Introduction}\label{sec:introduction}

\highlight{In modern astrophysics, observations across the entire electromagnetic spectrum ranging from radio and microwave to infrared, optical, ultraviolet, X-rays, and gamma-rays play crucial roles in building a comprehensive understanding of the Universe. Each regime probes distinct physical regimes, offering complementary perspective on astrophysical objects. While historically associated with cold, low-energy processes, radio observations now play a central role in exploring some of the Universe's most extreme and energetic phenomena -- including pulsars, supernova remnants, and fast radio bursts. The development of radio interferometry, especially Very Long Baseline Interferometry (VLBI), further enables extremely high-resolution and sensitive observations by combining signals from radio telescopes separated by large distances. VLBI has since been able to confirm various astrophysical phenomena that were once only predicted by theories e.g. event horizon of supermassive black holes \citep{EHT2019}, and spiral arms in accretion disk of massive protostars \citep{Burns2023}.}

Radio frequencies are divided into multiple frequency bands \citep[See Section 5.3.1;][]{ITUnsm}, and a portion of the bands, namely the low frequency (LF), medium frequency (MF), high frequency (HF), and very high frequency (VHF), span over 30 kHz -- 300 MHz, are crucial for various sciences in radio astronomy, including solar activity monitoring (solar burst and solar flares) and space weather observations, deep extragalactic observations of active galactic nuclei (AGNs), galaxies, and galaxy clusters, high-redshift observations, cosmic ray observations, and pulsar monitoring. Many instruments have been deployed to harness these frequencies\footnote{Although the International Telecommunication Union (ITU) defines this frequency range as low- to very-high frequencies, they are commonly referred to as \textit{low frequencies} in astronomy and astrophysics. Therefore, throughout the paper, we adopt the latter definition unless explicitly mentioned.}, mainly interferometers such as the Very Large Array \citep[VLA; 300 - 350 MHz and 74 MHz;][]{Thompson1980, Kassim1993}, the Low-Frequency Array \citep[LOFAR; 10 --240 MHz;][]{vanHaarlem2013}, the 21 CentiMeter Array \citep[21CMA; 50 -- 200 MHz;][]{Zhao2022}, the Murchison Widefield Array \citep[MWA; 70 -- 300 MHz;][]{Lonsdale2009}, \highlight{the Giant Metrewave Radio Telescope \citep[GMRT; 150 -- 1450 MHz;][]{Gupta2017}}, and the upcoming Square Kilometre Array Low Frequency Telescope \citep[SKA-low; 50 -- 350 MHz;][]{Labate2022}.

However, astrophysical sources are usually masked over by terrestrial radio signals (e.g. telecommunications) due to their significantly lower power. These terrestrial signals are referred to as radio frequency interference (RFI) \highlight{and pose a danger to radio astronomical observations. Although RFI can occur across the entire radio spectrum, it is particularly prevalent and problematic at lower frequencies \citep{Bentum2011}}. One solution is to build radio telescopes in radio quiet zones (RQZ) and radio notification zones (RNZ), but in countries with a majority of high and medium population density such as Malaysia, RFI evaluation becomes an extremely crucial procedure \citep{Abidin2012, Abidin2021}. \highlight{In addition to terrestrial sources, another emerging concern is the unintended interference from satellite constellations in low Earth orbit (LEO) e.g. Starlink. These satellites have been seen to affect frequencies used for radio astronomy, and its absence in RQZ regulations means that coexistence strategies are required even in remote or protected regions \citep{DiVruno2023}.}

This paper aims to investigate low frequency RFI in two locations within Malaysia; each representing an urban and rural environments. Malaysia's location close to the equator would allow for a wide sky coverage in both celestial hemispheres, which is currently covered by few, if not any radio telescope(s). Its central location in Southeast Asia would also significantly benefit VLBI experiments by serving as a vital link between the East Asian VLBI Network (EAVN; comprising of China, Korea, and Japan) and the Long Baseline Array (LBA; i.e. Australian telescopes/arrays), and even the European VLBI Network (EVN). This connection would enhance the overall network’s performance, particularly by improving short baselines and increasing overall sensitivity. Ultimately, alongside Thailand, Indonesia and Vietnam, involvement of Malaysia would further develop the goal of the regional South East Asian VLBI Network \citep[SEAVN;][]{Sugiyama2022}.

We organise this paper as follows: Section~\ref{sec:rfimonitoring} serves as a mini review of previous RFI monitoring that have been performed in Malaysia. Section~\ref{sec:gsk} summarizes the Generalized Spectral Kurtosis method which is used here for RFI detection. Section~\ref{sec:methodology} describes the instrumentation, observational parameters, and data analysis procedures. The results of the RFI monitoring are then presented in Section~\ref{sec:results}, along with associations of the RFI detected with spectrum allocations, potential origins, and presence of physical phenomena. Finally, we present and test a few potential RFI mitigation techniques in Section~\ref{sec:mitigation}.

\section{Previous RFI Monitoring Works in Malaysia}\label{sec:rfimonitoring}

RFI monitoring in Malaysia began as a pilot project for the establishment of radio astronomy research in the country. Naturally, these projects were performed in/near research institutes which are mainly located in dense populations such as in Universiti Malaya, Universiti Sultan Zainal Abidin, Universiti Teknologi MARA, Universiti Pendidikan Sultan Idris, and Universiti Kebangsaan Malaysia \citep{Abidin2009, Hamidi2011, Hamidi2012a, Umar2015, Zavvari2015}. As expected, most of these locations suffer from very high RFI, mainly due to mobile telecommunication, broadcasting (radio and television) and satellite communications \citep{Abidin2009, Abidin2010, Hamidi2011}. These RFI measurements targeted wide bands of up to 2.8 GHz, covering a number of important radio astronomy bands as shown in Table \ref{tab:frequencybands}.

\begin{table}
    \centering
    \caption{Important frequency bands used in radio astronomy below 2.8 GHz \highlight{\citep[Sources:][]{Abidin2009, ITUhra}.}}
    \label{tab:frequencybands}
    \begin{tabular}{cl}
        \hline
        Frequency range (MHz) & Radio astronomy usage \\
        \hline
        13.36 - 13.41 & Solar observation \\
        25.55 - 25.67 & Jupiter observation \\
        37.50 - 38.25 & Continuum observation \\
        73.00 - 74.60 & Solar wind measurement \\
        150.05 - 153.00 & Pulsar monitoring \\
        322.00 - 328.85 & Deuterium observation \\
        406.00 - 410.00 & Pulsar monitoring \\
        608.00 - 614.00 & Continuum and interferometry \\
        1400.00 - 1427.00 & 21-cm hydrogen observation \\
        \highlight{1610.60 - 1613.80} & \highlight{Hydroxyl observation} \\
        \highlight{1660.00 - 1670.00} & \highlight{Hydroxyl observation} \\
        \highlight{1718.80 - 1722.20} & \highlight{Hydroxyl observation} \\
        2655.00 - 2700.00 & Pulsar monitoring and interferometry \\
        \hline
    \end{tabular}
\end{table}

Further research by \cite{Abidin2013} and \cite{Umar2014} documented the influence of population density on RFI levels. Correlation between the noise levels and population density was seen, largely due to the heavy contamination from telecommunication services in urban areas. Emissions of these sources typically appear between 900 - 1200 MHz and 1600 - 2000 MHz \citep{Abidin2009, Hamidi2011, Hamidi2012a, Hamidi2012b}. This suggests that radio telescopes should ideally be built in sparsely populated regions, with a suggested population density threshold of 150 people per square kilometre\highlight{\footnote{\highlight{Depending on the advancement of technology and its prevalence in the common household, the population density threshold will vary across time.}}} \citep{Umar2014}.

Following that, various rural areas in Malaysia \citep[e.g. Jelebu, Sekayu, Langkawi, Behrang, Merang;][]{Abidin2010, Abidin2011, Umar2012, Noorazlan2013, Zafar2017, Shafie2017, Zafar2018, Shafie2021} were explored and notably lower RFI levels were detected, with exceptions in satellite and broadcasting frequency bands. Observations in narrow spectral windows (e.g. deuterium, 21-cm hydrogen, hydroxyl) reveal mixed results -- few locations show absence of RFI in certain bands \citep[e.g. deuterium;][]{Zavvari2015}, while others display RFI spilling into the vicinity of important bands \citep[e.g. hydroxyl;][]{Zavvari2015} or even contaminating the band itself (e.g. deuterium; \citealt{Abidin2012}, and L-band; \citealt{Abidin2013, Zafar2017}).

\begin{figure*}
    \centering
    \includegraphics[width=0.95\linewidth]{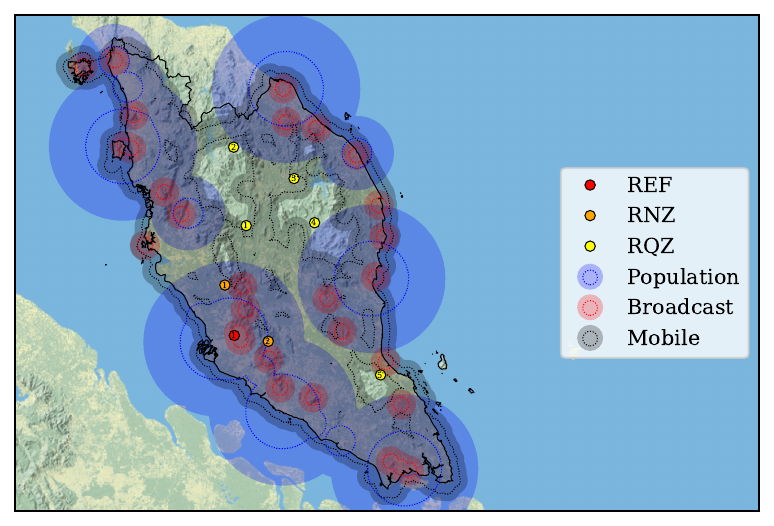}
    \caption{Identification of possible RNZs and RQZs in Peninsular Malaysia based on the minimum weightage method as described in \citealt{Abidin2021}. The coloured semi-transparent areas denote the RQZ-defined contamination range, while the dotted lines within them denote the RNZ-defined contamination range, each for population densities displayed in blue, or for each type of RFI source i.e. \highlight{broadcasting stations (TV, AM, and FM) in red, and mobile telecommunication (2G, 3G, and 4G) in grey}. The red, orange, and yellow points indicate the locations of reference (1: Universiti Malaya), candidate RNZ (1: Universiti Pendidikan Sultan Idris, 2: Jelebu), and candidate RQZ (1: Kuala Medang, 2: Grik, 3: Gua Musang, 4: Ulu Tembeling, 5: Selendang) sites respectively i.e. where the RFI monitorings were performed in \citealt{Abidin2021}. Note that the population data, and mobile and broadcast station locations have been updated to the latest available data, and this may have redacted the RNZ and RQZ status of a few sites (namely Universiti Pendidikan Sultan Idris, Gua Musang, and Selendang).}
    \label{fig:rqz-map}
\end{figure*}

The current most comprehensive RFI study conducted in Peninsular Malaysia aimed to investigate radio quiet zones (RQZ) and radio notification zones (RNZ). \cite{Abidin2021} stated that although RQZs are ideal locations for radio astronomy, they are both logistically impractical and almost impossible to realize in a dense population like Malaysia. RNZs become the next best option where there are no authoritative restrictions, but shared frequencies are \textit{notified} to the corresponding parties. They listed five RQZ and two RNZ candidates (with Universiti Malaya as a reference) based on average monthly humidity/rainfall, population density, and distance from mobile and broadcasting stations using the minimum weightage method. Shown in Figure \ref{fig:rqz-map} is a map of the candidate RNZs and RQZs with the minimum weightage boundaries. When RFI monitoring was performed up to 4 GHz, they found that the average noise in all the sites are below the International Telecommunication Union (ITU) recommendation for continuum and spectral line astronomy observations\footnote{ITU-R RA.769-2: Protection criteria used for radio astronomical measurements.}, with Ulu Tembeling and Jelebu regarded as the best RQZ and RNZ sites respectively.

However, most of these measurements were performed with various limitations. Firstly, most observations cover very wide frequency ranges of up to 9 GHz, resulting in a large band separations which may not capture narrow band RFI \citep[e.g.][]{Abidin2012}. Narrow band RFI measurements in contrast target frequencies above VHF VLBI windows or specific spectral lines. Basic and small antennas (9" omnidirectional, 5.2 cm discones, and 50 cm Yagi antennas) which were used may also miss weaker RFI signals that would otherwise be captured by sensitive radio telescopes \citep[e.g.][]{Noorazlan2013}. In addition, the RFI data were taken over large cadences and averaged, meaning that transient RFI emissions cannot be studied (except in \citealt{Abidin2011}, but no detailed results were discussed). Finally, majority of the studies perform little to no statistical evaluations (e.g. only using noise average and standard deviations) which may not manifest the true nature of the RFI. Only in a few cases \citep[e.g.][]{Abidin2011, Noorazlan2013, Zafar2017}, higher order Kurtosis statistics were utilized, but exclusively in the temporal domain and were not discussed in detail.

\section{Generalized Spectral Kurtosis (GSK)}\label{sec:gsk}

\cite{Nita2007} proposed the spectral kurtosis (SK) estimator method to detect (and mitigate) RFI in real time. The generalized spectral kurtosis (GSK) was then introduced by \cite{Nita2010a}, a method that interposes the time-averaging factor to address hardware limitations of standard correlators. Generalizing the shape factor in the gamma distribution obeyed by the power spectrum allows it to be applied to instruments which can only output already averaged data. These methods have been applied to various instruments with noteworthy performance e.g. the Korean Solar Radio Burst Locator \citep[KSRBL;][]{Dou2009}, the Expanded Owens Valley Solar Array \citep[EOVSA;][]{Nita2016}, the Canadian Hydrogen Intensity Mapping Experiment instrument \citep[CHIME;][]{Taylor2019}, and the Parkes Telescope \citep{Wongphechauxsorn2023}. This method has also been evaluated to be effective for cross-correlation application, potentially applicable in VLBI \citep{Nita2020}. SK's simplicity further allows for real-time calculations and application on radio receivers using low-cost FPGAs \citep{QuirsOlozbal2016, Taylor2019}.

As described by \cite{Nita2010a}, the GSK estimator is defined by
\begin{equation}
    \widehat{SK} = \frac{MNd + 1}{M - 1} \left(\frac{M S_2}{S_1} - 1\right) \label{eq:gske}
\end{equation}
where $S_1 = \sum^M_{i=1} (\sum^N_{j=1} x_j)_i$, $S_2 = \sum^M_{i=1} (\sum^N_{j=1} x_j)^2_i$, and $M$, $N$, and $d$ are the accumulation length of power spectral density $x_i$, number of averaged spectra prior to recording, and shape factor (explained later in Section \ref{sub:modifiedgsk}) respectively. The mean and first central moments of the pdf of $\widehat{SK}$ can then be analytically defined based on Equation \ref{eq:gske} \citep[refer][]{Nita2010b} as 
\begin{align*}
    \mu'_1 &= E(\widehat{SK}) \equiv 1\\
    \mu_2 &= \frac{2Nd(Nd + 1) M^2 \Gamma(MNd + 2)}{(M - 1)\Gamma(MNd + 4)}\\
    \mu_3 &= \frac{8Nd(Nd + 1) M^3 \Gamma(MNd + 2)}{(M - 1)^2 \Gamma(MNd + 6)}\\
            & \times [(Nd + 4)MNd - 5Nd - 2]\\
    \mu_4 &= \frac{12Nd(Nd + 1) M^4 \Gamma(MNd + 2)}{(M - 1)^3 \Gamma(MNd + 8)}\\
            & \times (M^3N^4d^4 + 3M^2N^4d^4 + M^3N^3d^3 + 68M^2N^3d^3\\
            & - 93MN^3d^3 + 125M^2N^2d^2 - 245MN^2d^2\\
            & + 84N^2d^2 - 32MNd + 48Nd + 24).
\end{align*}
where $\Gamma$ is the gamma function. Notice that the GSK estimator has a mean of 1, and this eliminates the need for the evaluation of the mean background level which can be skewed in the presence of strong RFI.

To approximate the SK distribution, we utilize the Pearson Type III probability distribution function (pdf), which follows the generally asymmetrical frequency curve of background radio emissions in nature \citep{Pearson1895}. The Type III cumulative function (CF) is consequently defined as
\begin{equation}
    CF(\xi, \alpha, \beta, \delta) = \Gamma_x \left(\beta, \frac{\xi - \delta}{\alpha}\right) \bigg/ \Gamma(\beta)
\end{equation}
where $\alpha = \mu_3/2\mu_2$, $\beta = 4\mu^3_2/\mu^2_3$, $\delta = 1 - 2\mu^2_2$, and $\Gamma_x(\beta, x) = \int_0^x t^{\beta - 1}e^{-t}\,\mathrm{d}t$ is the incomplete gamma function. \highlight{This approximation is accurate up to the third order, especially for large accumulation lengths $M > 1000$ \citep{Nita2010a}.} By setting a predefined false alarm probability $P_{\mathrm{FA}}$ (usually $0.13499\%$ which is equivalent to the $P_{\mathrm{FA}}$ of a $3\sigma$ threshold applied to a Gaussian distribution), one can ultimately obtain the bounds of good $\widehat{SK}$ (denoted as $\delta_{\rm lower}$ and $\delta_{\rm upper}$) by optimizing the $P_{\mathrm{FA}}$ with the Type III CF and complementary CF, and then flag the respective channels in the data accordingly. 

\section{Methodology}\label{sec:methodology}

\subsection{Instruments}\label{sub:instruments}

Our observational setup consists of two main instruments -- A dual-polarization logarithmic periodic antenna and a Liquid Instruments Moku:Lab. 

The dual-polarization logarithmic periodic antenna consists of two log-periodic antennas arranged perpendicularly along their principle axis, allowing it to measure two linear polarizations at once (horizontal and vertical). The antenna has a total length of 180 cm and a width of 202 cm, equipped with a total of 17 dipoles with a standing wave ratio of 2.5. The antenna operates optimally up to 1000 MHz, with a typical gain of 6.5 dBi and input impedance of $50~\Omega$. At 100 MHz, the antenna has a beamwidth of $118.6^{\circ}$ and $60.4^{\circ}$ in the H- and E-planes respectively.

The Liquid Instruments Moku:Lab is a reconfigurable hardware platform mainly controlled by Field Programmable Gate Arrays (FPGA), and equipped with multiple low-noise analogue inputs and outputs \citep{Mokulab}. In our case, the spectrum analyser mode was used to measure the radio signals from the antenna. Table \ref{tab:mokuspecs} shows the specifications of the Moku:Lab in the spectrum analyser mode. The instrument's input noise follows the following profile (for $f$ in kHz): 
\begin{equation}
    S_{\eta}(f) \approx 13.9~\mathrm{nV} \times \sqrt{1 + \frac{220~\mathrm{kHz}}{f}}
\end{equation}
For our observations between $0 - 250~\mathrm{MHz}$, the average input noise is approximately $13.6~\mathrm{nV}$, \highlight{with a noise factor of 2 dB (equivalent to a noise temperature of $\sim 170~\mathrm{K}$).}

\begin{table}
    \centering
    \caption{Specifications of Moku:Lab in spectrum analyser mode.}
    \label{tab:mokuspecs}
    \begin{tabular}{cc}
        \hline
        Parameter & Value \\
        \hline
        Frequency range & DC to $250~\mathrm{MHz}$ \\
        Frequency channels & $1023$ \\
        Channel spacing & $244140.62~\mathrm{Hz}$ \\
        Impedance (AC) & $50~\mathrm{\Omega}$ \\
        Attenuation (AC) & $0~\mathrm{dB}$ \\
        Voltage range & $\pm 0.5~\mathrm{V}$ \\
        Voltage sensitivity & $-130~\mathrm{dBm}$ \\
        \hline
    \end{tabular}
\end{table}

\subsection{Observations}\label{sub:observations}

\begin{figure}
    \centering
    \includegraphics[width=0.95\linewidth]{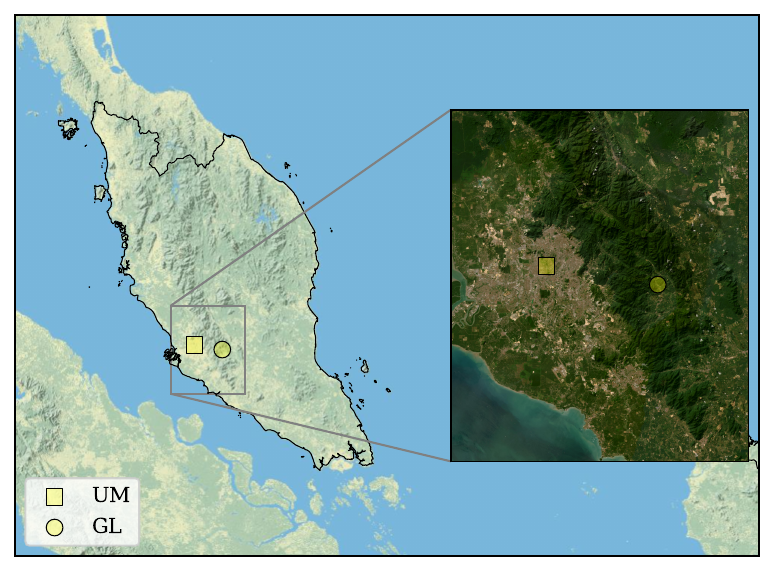}
    \caption{A map of Peninsular Malaysia with the two radio observation stations. The main map represents the Natural Earth physical map, while the inset shows a zoomed-in satellite image around the two stations, depicting their respective urban and rural environments. The GL station can also be seen to be surrounded by the Titiwangsa Mountains which provides shielding from urban radio signals.}
    \label{fig:map}
\end{figure}

\begin{figure}
    \centering
    \includegraphics[width=0.95\linewidth]{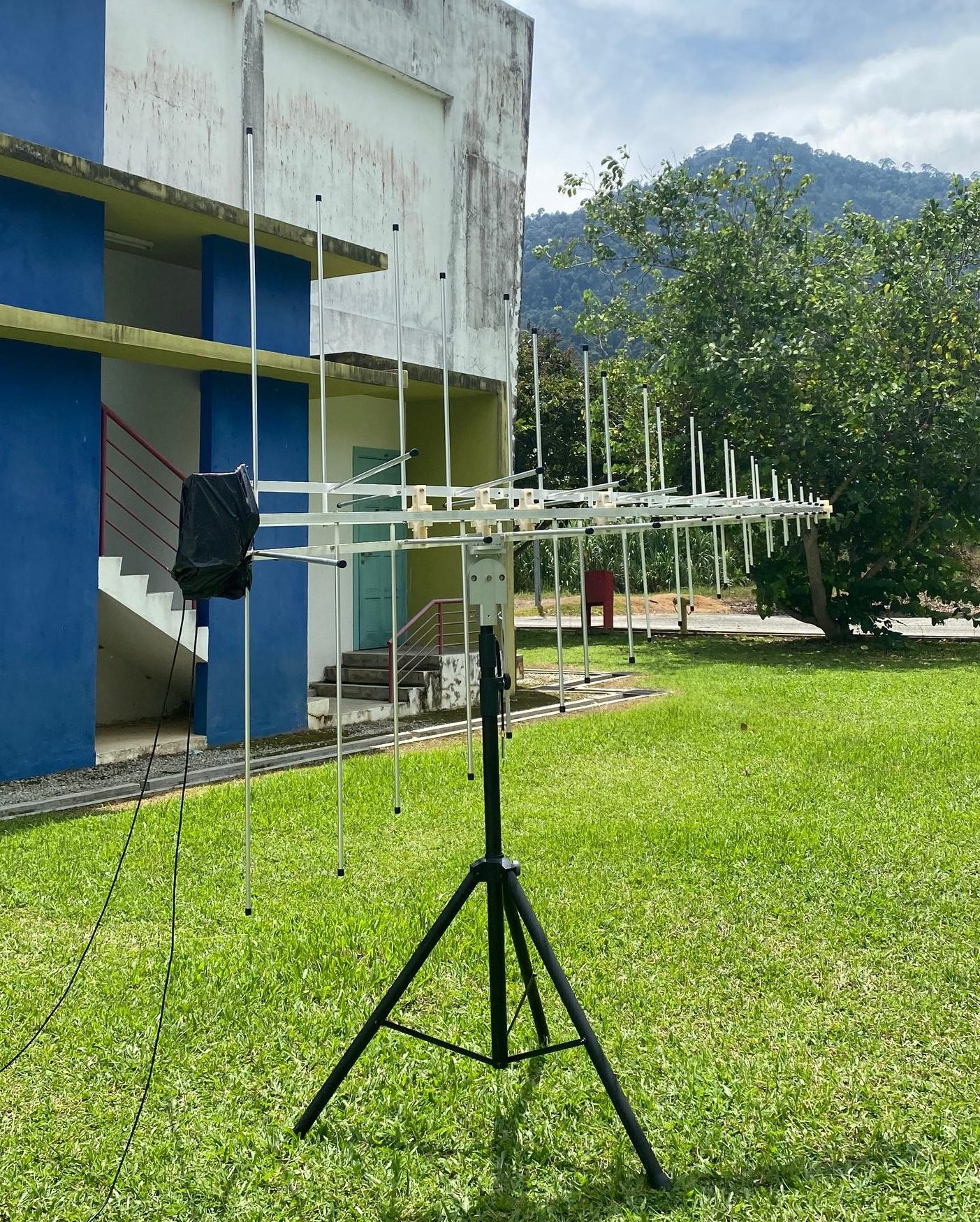}
    \caption{Side-view of the dual-polarization logarithmic periodic antenna set up in GL for radio observation. The two coaxial cables connected to the base of the antenna are connected to the spectrum analyser.}
    \label{fig:antenna-setup}
\end{figure}

The radio observations were done in two locations. The first location is the Physics Department, Universiti Malaya (UM), located in Malaysia's capital Kuala Lumpur, Malaysia ($3^{\circ}7'23.754''\mathrm{N}$, $101^{\circ}39'10.6812''\mathrm{E}$, 78 m altitude). UM represents a highly populated urban area (Population density: $8235~\mathrm{km}^{-2}$), prone to be contaminated by large amounts of RFI. The second location is the Glami Lemi Biotechnology Research Centre (GL), located in Jelebu, Negeri Sembilan, Malaysia ($3^{\circ}3'12.6216''\mathrm{N}$, $102^{\circ}3'51.7032''\mathrm{E}$, 144 m altitude). GL has been suggested as a RNZ due to its low population density (Population density: $35.11~\mathrm{km}^{-2}$) and shielding from surrounding hills, and combined with the location's low average humidity and precipitation makes it a prominent location for radio astronomy observations \citep{Abidin2021}. The two locations are marked in the map shown in Figure \ref{fig:map}. \highlight{The antenna was azimuthally oriented to point away from the largest nearby structure(s) (e.g. buildings, hills/mountains, etc.) to maximize the capture of RFI present in the environment.} The observation schedule for both locations are shown in Table \ref{tab:schedule}, and the radio spectrum was obtained approximately every 0.3 seconds. Figure \ref{fig:antenna-setup} shows a side-view of the antenna deployed in one of the observation locations.

\begin{table}
    \centering
    \caption{Observation schedule. Start and end times are written in GMT+8 (Malaysian Time).}
    \label{tab:schedule}
    \begin{tabular}{ccccc}
        \hline
        Location & Start Date & End Date & Start Time & End Time \\
        \hline
        UM & 2 Nov 2023 & 3 Nov 2023 & 06:08:06 & 06:05:11 \\
        GL & 24 Oct 2023 & 25 Oct 2023 & 16:01:05 & 17:00:58 \\
        \hline
    \end{tabular}
\end{table}

\subsection{Modified GSK Estimator}\label{sub:modifiedgsk}

To detect frequency channels contaminated by RFI, we adopt the generalized spectral kurtosis (GSK) method as explained in Section \ref{sec:gsk}. In this work, since the value of $N$ for our instrument is not precisely known, we assume $N$ to have a value of unity, representing unaveraged power spectral density \citep{Smith2022}. 

Equation \ref{eq:gske} indicates that the $\widehat{SK}$ spectra will typically appear flat, with the exception of RFI-contaminated frequencies. However, due to variations in instrument characteristics \citep[e.g. digitizer quantization;][]{Nita2020}, the $\widehat{SK}$ spectra will not be normalized to unity. To address this, we could normalize the $\widehat{SK}$ spectra by calculating its mean. To be exact, the spectra is normalized using the shape factor \citep{Nita2020} which is defined based on the ideal case of Equation \ref{eq:gske} as
\begin{equation}\label{eq:shapefactor}
    d = \frac{M - \mu + 1}{M \mu}
\end{equation}
where $\mu = \left\langle \widehat{SK} (d \equiv 1) \right\rangle$ i.e. the mean of $\widehat{SK}$. This empirical approach means that the true value of $N$ isn't required as its value will be influenced by the $d$ estimates.

We then propose the usage of the median, $\hat{\mu}$ instead of the mean to calculate the shape factor, modifying Equation \ref{eq:shapefactor} into
\begin{equation}
    d = \frac{M - \hat{\mu} + 1}{M \hat{\mu}}
\end{equation}
This helps to safeguard the integrity of $\widehat{SK}$ from the influence of significantly RFI-contaminated frequencies, assuming that there are more uncontaminated frequencies than contaminated ones. 

Similar to the EOVSA \citep{Nita2016}, the dual-polarization data are flagged separately and then combined using an OR operator. This is done to capture RFIs in each channel and flag them in both linear polarizations without any bias, though this does increase the expected $P_{\mathrm{FA}}$ by a factor of $(2 - P_{\mathrm{FA}})$, for which in the case of the Gaussian $3\sigma$ threshold, to $0.270\%$ \citep{Nita2020}.

\section{Results and Discussion}\label{sec:results}

\subsection{Urban vs Rural RFI}\label{sub:urbanruralrfi}

\begin{figure}
    \centering
    \includegraphics[width=0.75\linewidth]{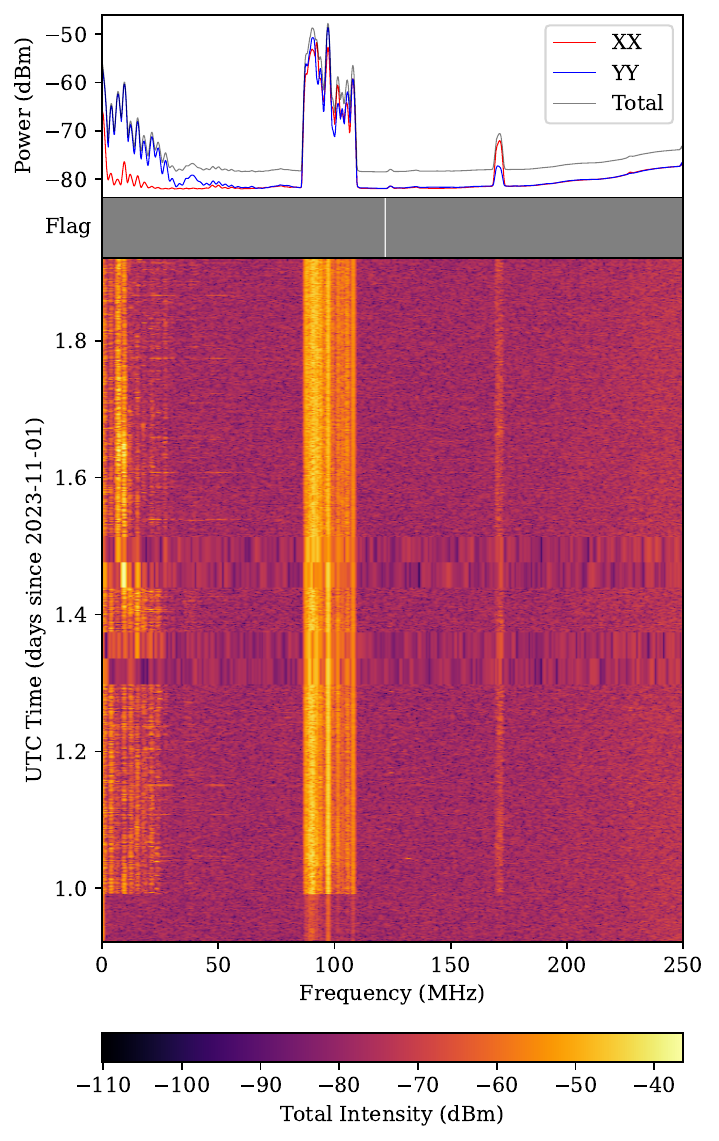}
    \caption{Overall result plot for UM data. The top panel shows the time-averaged power spectra for both linear polarizations and the total intensity, the middle panel shows the SK flags (gray lines denote the flagged frequencies), and the bottom panel shows a waterfall plot of the total intensity for the entire observation period.}
    \label{fig:plot-um}
\end{figure}

\begin{figure}
    \centering
    \includegraphics[width=0.75\linewidth]{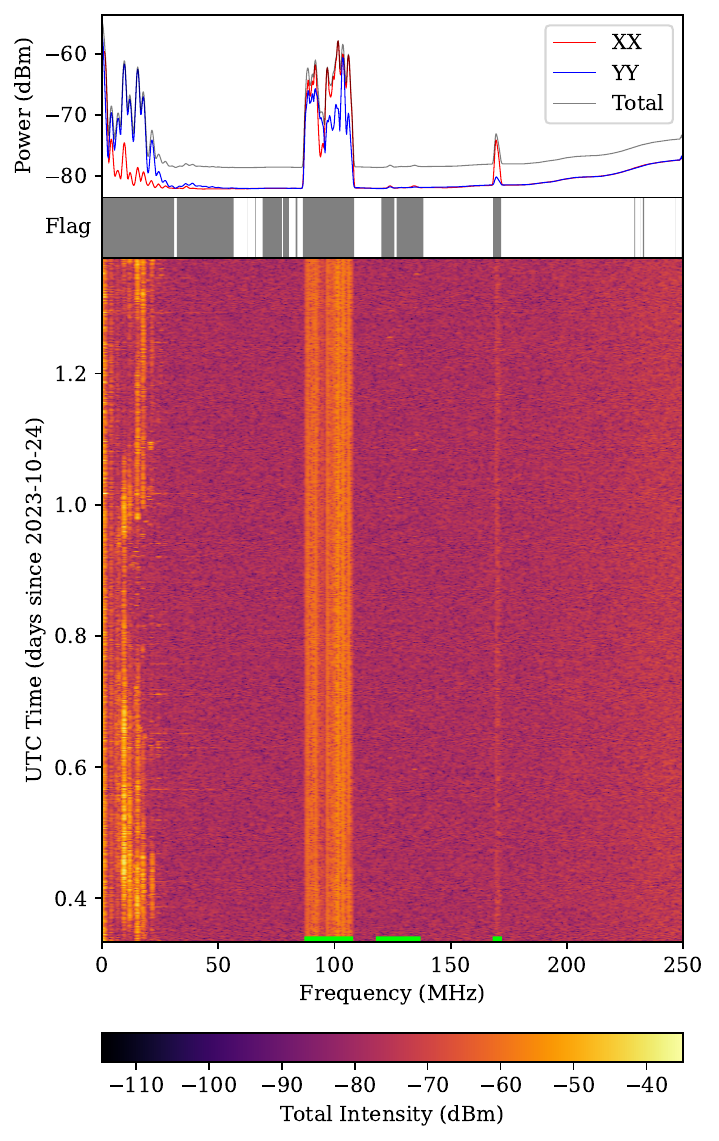}
    \caption{Similar plot to Figure \ref{fig:plot-um} but for GL data. \highlight{The green markers at the bottom of the waterfall plot depicts the three particular groups of narrowband RFI detected; (from left to right) FM analogue radio broadcast (87 - 108 MHz), aeronautical mobile or airband (117.825 - 137 MHz), and Malaysian Government mobile services (162.0375 - 174 MHz).}}
    \label{fig:plot-gl}
\end{figure}

\begin{figure}
    \centering
    \includegraphics[width=0.95\linewidth]{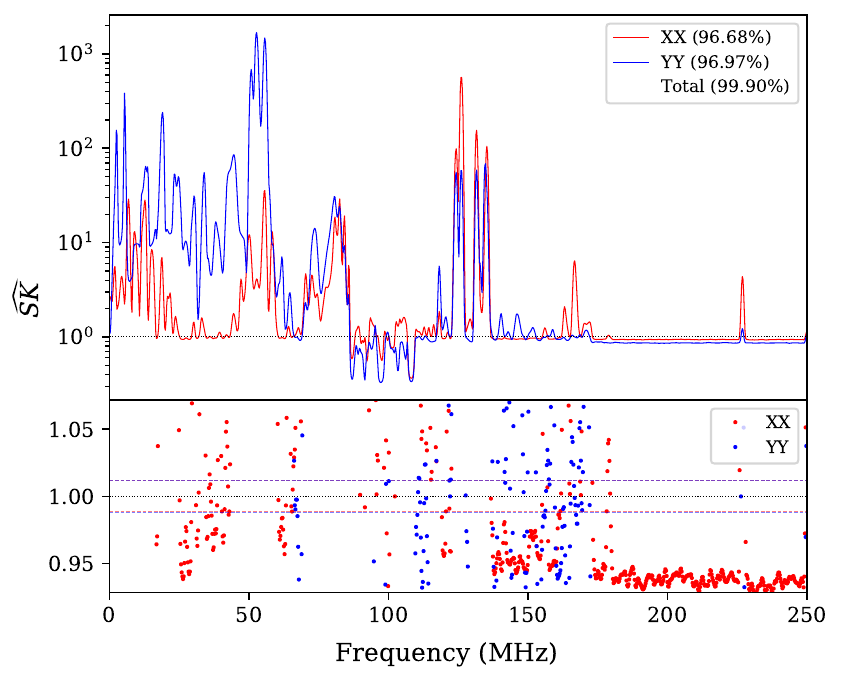}
    \caption{SK spectra in both linear polarizations for UM data. The top panels show the entire range of SK values while the bottom panels show SK values within the vicinity of the respective SK lower and upper bounds. The percentages associated to each polarization (written in the legend) corresponds to their respective flagging fractions.}
    \label{fig:sk-um}
\end{figure}

\begin{figure}
    \centering
    \includegraphics[width=0.95\linewidth]{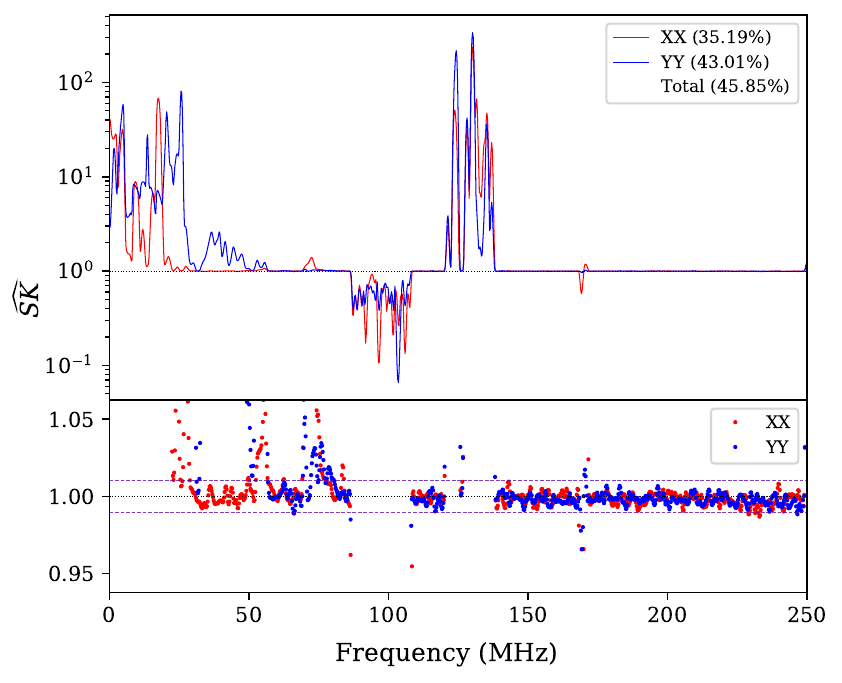}
    \caption{Similar plot to Figure \ref{fig:sk-um} but for GL data.}
    \label{fig:sk-gl}
\end{figure}

Figures \ref{fig:plot-um} and \ref{fig:plot-gl} show the overall result plots for UM and GL data respectively, with the middle panel showing the frequency channels flagged using the modified SK (shown as grey bars). Figures \ref{fig:sk-um} and \ref{fig:sk-gl} show the SK spectra for UM and GL data respectively. For the UM data, 1022 channels out of 1023 channels (99.90\%) are flagged, whereas for the GL data, 469 channels out of 1023 channels (45.85\%) are flagged.

\highlight{Table \ref{tab:range} shows the power ranges of the averaged spectra, while Table \ref{tab:results} shows the summary for SK statistics, both for UM and GL data.} The averaged power spectra range and statistics (minimum, maximum, mean, standard deviation, and median) in the UM data consistently surpasses that of the GL data across all cases, with the median SK values for the UM data in both polarizations also exceeding those of the GL data.

These results indicate a higher baseline in the power spectrum and a larger degree of RFI contamination in urban areas, as represented by UM, compared to rural areas, represented by GL. This similar to the results of \cite{Umar2013, Abidin2021} where RFI levels were seen to depend strongly on population density, primarily attributed to the presence of more telecommunication and broadcasting stations. However, this was previously reported to occur at higher frequencies \citep[e.g. near 1000 and 1800 MHz;][]{Abidin2009}, and our results demonstrate that it does affect the LF - VHF range as well. 

\highlight{In RFI-dominated spectra, the SK shape factor becomes biased, increasing false positive flags \citep{Nita2016b}, which may have been the case for the UM data. Nonetheless, several other factors can physically explain the near-complete RFI contamination in urban environments. Strong narrowband signals can leak into adjacent channels during digitization, leading to overflagging by spectral kurtosis \citep{Smith2022}. Densely packed emitters and poorly tuned devices produce overlapping sidebands that mimic broadband interference. Transient emissions from electronic systems and intermodulation from non-linear components further add to the noise. Additionally, urban reflections and multipath propagation broaden and intensify RFI signatures \citep{Parsons1989}. Combined, these effects make narrowband sources appear as continuous interference, especially in densely populated areas.}

\begin{table}
    \centering
    \caption{Averaged power spectra range and basic statistics in the UM and GL data. All values are in dBm.}
    \label{tab:range}
    \begin{tabular}{ccccccc}
        \hline
        \multirow{2}{*}{Statistic} & \multicolumn{3}{c}{UM} & \multicolumn{3}{c}{GL}\\
        & XX & YY & Total & XX & YY & Total \\
        \hline
        Minimum & -82.0518 & -81.9125 & -114.7707 & -82.1486 & -82.0870 & -114.7737 \\
        Maximum & -51.7386 & -48.5810 & -24.0870 & -57.8832 & -57.3436 & -24.5401 \\
        Mean & -78.8963 & -77.6990 & -74.5701 & -79.4308 & -79.0911 & -75.6161 \\
        Std. Dev. & 6.1742 & 6.8550 & 7.8493 & 4.9507 & 4.6944 & 6.2942 \\
        Median & -81.4583 & -80.8075 & -76.1759 & -81.5000 & -81.4369 & -76.5144 \\
        \hline
    \end{tabular}
\end{table}

\begin{table}
    \centering
    \caption{Summarized SK statistics for UM and GL data in both linear polarizations. $\hat{\mu}$ is the median $\widehat{SK}$ value, $d$ is the shape factor, $\delta$ are the upper and lower $\widehat{SK}$ thresholds, and $N$ are the flagged channel counts (and fractions).}
    \label{tab:results}
    \begin{tabular}{cccccc}
        \hline
        \multirow{2}{*}{Statistic} & \multicolumn{2}{c}{UM} & \multicolumn{2}{c}{GL} \\
        & XX & YY & XX & YY \\
        \hline
        $\hat{\mu}$ & 1.079763 & 1.171981 & 1.013884 & 1.013477 \\
        $d$ & 0.926057 & 0.853112 & 0.986293 & 0.986690 \\
        $\delta_{\rm lower}$ & 0.988417 & 0.988168 & 0.989680 & 0.989682 \\
        $\delta_{\rm upper}$ & 1.011790 & 1.012054 & 1.010480 & 1.010479 \\
        $N_{\rm lower}$ & 493 (48.19\%) & 492 (48.09\%) & 105 (10.26\%) & 97 (9.48\%) \\
        $N_{\rm upper}$ & 496 (48.48\%) & 500 (48.88\%) & 255 (24.93\%) & 343 (33.53\%) \\
        $N_{\rm subtotal}$ & 989 (96.67\%) & 992 (96.97\%) & 360 (35.19\%) & 440 (43.01\%) \\
        $N_{\rm total}$ & \multicolumn{2}{c}{1022 (99.90\%)} & \multicolumn{2}{c}{469 (45.85\%)} \\ 
        \hline
    \end{tabular}
\end{table}

\subsection{Spectrum Allocation}\label{sub:spectrumallocation}

As the UM spectrum is flagged in almost all frequency channels, only the GL spectrum is used to characterize specific RFI sources. However, it is noteworthy to mention that both the UM and GL data possess the same RFI-channel signature. There are two general types of RFI detected in the data: persistent RFI (with duty cycles of $\gtrsim 50\%$) and transient RFI (with duty cycles of $\lesssim 50\%$). The three persistent RFI signals detected in the data are from 0 to 31.25 MHz, between 86.43 and 108.40 MHz, and between 168.21 and 171.63 MHz (hereinafter referred to as RFI-P1, RFI-P2, and RFI-P3 respectively), while the two transient RFI signals detected in the data are between 32.23 and 56.64 MHz, and between 120.12 and 138.18 MHz (hereinafter referred to as RFI-T1 and RFI-T2 respectively). Any spectrum allocations mentioned are based on the ITU and the Malaysian Communications and Multimedia Commission (MCMC) based on the 2022 Spectrum Allocation document \citep{MCMCSP2022}.

\subsubsection{Persistent RFI with Diurnal Variation}

\begin{figure}
    \centering
    \includegraphics[width=0.75\linewidth]{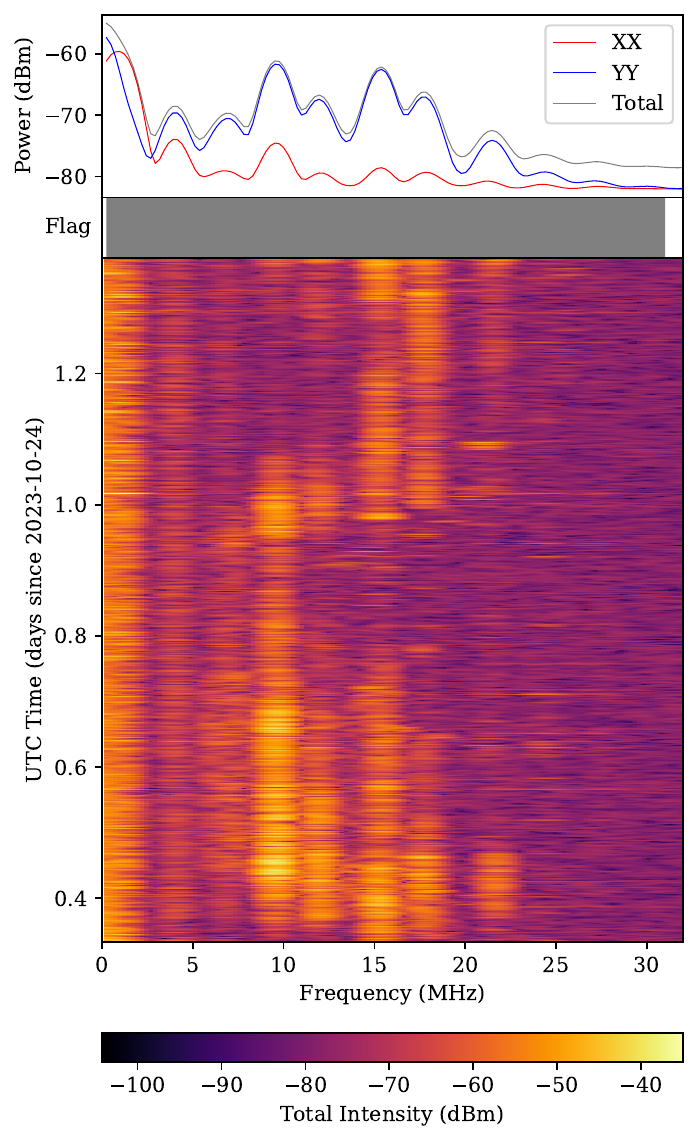}
    \caption{Similar plot to Figure \ref{fig:plot-um} but for GL data zoomed into the RFI-P1 frequency range.}
    \label{fig:plot-gl-rfi-p1}
\end{figure}

RFI-P1 (0 to 31.25 MHz), shown in Figure \ref{fig:plot-gl-rfi-p1}, has a unique structure consisting of two distinct emission patterns; a group with relatively consistent emission below 5 MHz and a group of signals with temporally varying amplitudes peaking at -24.54 dBm. Hereinafter, the groups are referred to as RFI-P1a and RFI-P1b respectively. The entire LF -- HF range is allocated for all types of operations including mobile services, broadcasting, maritime communications, and amateur use. However, the relatively wide channel spacing in our observation means that these narrow-band emissions are averaged in each frequency bin, retaining only large-scale patterns. 

At such low frequencies, RF1-P1a should consist of signals from local and regional radio sources propagating directly or via ground-reflected waves to the data acquisition locations \citep{Hufford1952, Hill1982, DeMinco2000}. This is reinforced with the consistency in the signal throughout the entire observation period, which can be attributed to the \highlight{maximal} propagation effects at this frequency range. 

\begin{figure}
    \centering
    \includegraphics[width=0.95\linewidth]{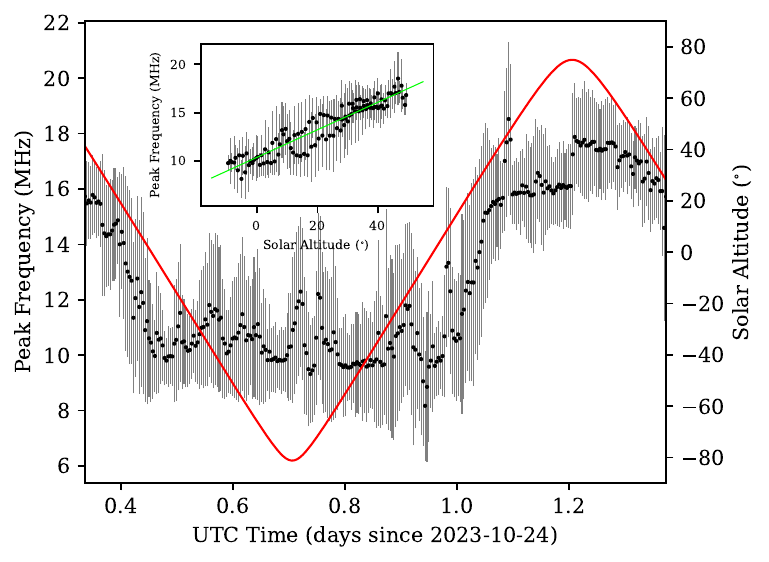}
    \caption{Summary plot of the diurnal pattern in \highlight{RF1-P1b} of the GL data. The main plot shows the evolution of the peak emission frequency \highlight{for emissions between 0 and 31.25 MHz}, with the black markers and grey error bars representing binned data over every 5-minute interval with their 1$\sigma$ respectively, and the red line representing the solar altitude. The inset shows the correlation fit between the peak emission frequency in the GL data and the solar attitude, with the green line representing a linear fit to the data. Note that only the data for solar altitudes between $-10^{\circ}$ and $50^{\circ}$ are considered for the fitting due to the constant darkness from dusk to dawn and the saturation of the formation of the ionospheric D layer near noon.}
    \label{fig:diurnal-plot}
\end{figure}

In contrast, RF1-P1b has a distinct signature in which the amplitudes in different channels varies with a diurnal pattern in the data as shown in Figure \ref{fig:diurnal-plot}. The frequency at which the brightest emission is observed at a certain time correlates linearly with the solar altitude with a correlation coefficient of 0.9392 as shown in the inset of Figure \ref{fig:diurnal-plot}. This diurnal variation in the peak frequency can be explained by skywave propagation of radio signals in the ionosphere, and is usually prevalent at frequencies below 35 MHz \citep{Wagner1995, Zhou2019}.

In daylight, Lyman-$\alpha$ and X-ray solar radiation enhances the formation of the ionospheric D layer, increasing the ionosphere's opacity of low-frequency signals \citep{Rumi1960, Bibl1961}. This enables higher frequencies to pass through the D layer and reflect off the ionospheric F layers. During the night, with no solar radiation, the D layer significantly thins (remaining due to cosmic rays), facilitating more efficient skywave propagation of low-frequency signals. \highlight{Small scale disturbances in the trend may be solar-induced e.g. solar X-ray flares enhancing D layer photoionization \citep{Thomson2001, Kumar2018, Rozhnoi2019, Briand2022, Shamsuddin2022, Gu2023} and/or of atmospheric origin e.g. thunderstorms causing atmospheric gravity waves (AGWs) and lightning-induced quasi-electrostatic fields which can disrupt the D layer \citep{Cheng2007, Lay2011, Shao2012, Redoblado2022}.}

\subsubsection{Other Persistent RFI}

RFI-P2 (86.43 to 108.40 MHz) exhibits two primary signal groups, delineated by a dip at 95 MHz. While the signal does feature smaller subband divisions, its temporal consistency is evident, with a maximum amplitude of -51.63 dBm. These bands are allocated to mobile services and broadcasting by the ITU \highlight{(87 - 108 MHz for Region 3)} and MCMC, but is mainly occupied by frequency modulation (FM) analogue radio broadcast. This RFI signal was previously seen in RFI monitoring in the same locations \citep{Abidin2011, Abidin2021}, and is consistent with monitoring performed in other locations in Malaysia \citep{Umar2015} as well as in neighbouring countries e.g. Thailand and Indonesia \citep{Hidayat2014, Jaroenjittichai2017}.

RFI-P3 (168.21 to 171.63 MHz) displays a distinctive single-band signature with a peak amplitude of -59.08 dBm. This signal falls within the mobile services allocation \highlight{by the ITU (162.0375 - 174 MHz in all regions)}, with the 169 to 173.50 MHz range allocated to the Malaysian Government (MLA14).

\subsubsection{Transient RFI}

RFI-T1 (32.23 to 56.64 MHz) is marked by random emissions dispersed in both time and frequency, reaching a peak intensity of -51.23 dBm. This frequency band has diverse allocations, encompassing mobile services, radiolocation, broadcasting, amateur operations, space research, and radio astronomy. Notably, the 5.149 ITU radio astronomy footnote designates protection for the 37.5 to 38.25 MHz band, specifically for Jupiter's decametric radiation observations. This serves as a reminder to conduct thorough RFI assessments even during observations within radio astronomy-protected bands.

\begin{figure}
    \centering
    \includegraphics[width=0.75\linewidth]{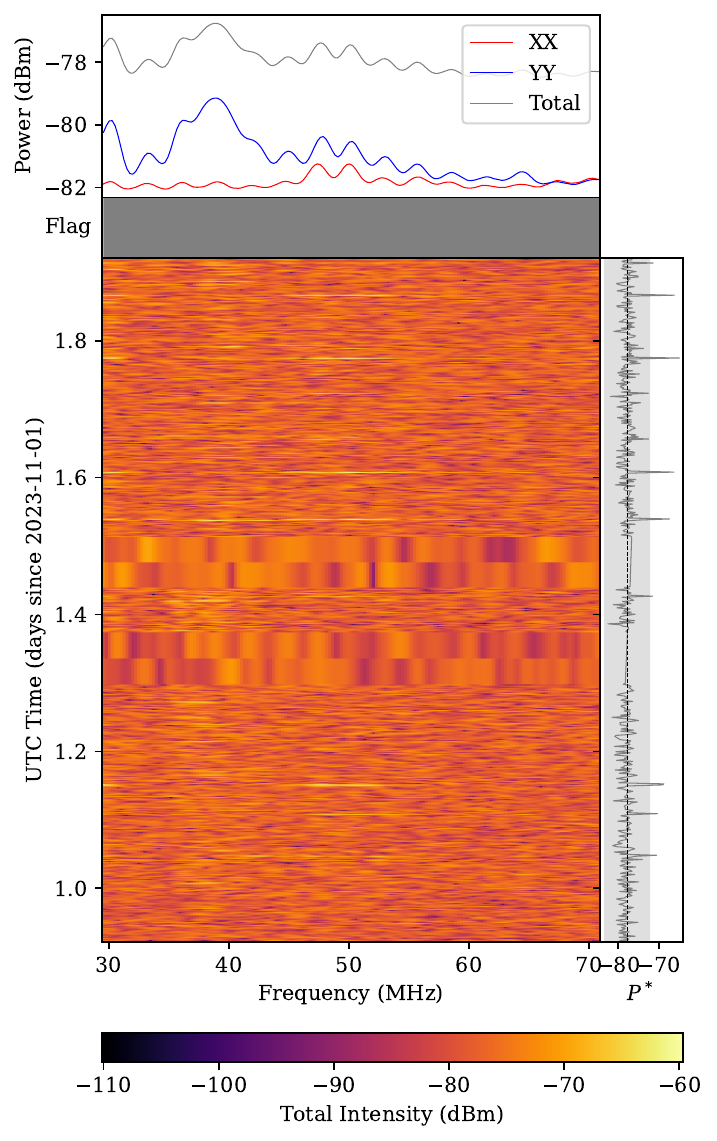}
    \caption{Similar plot to Figure \ref{fig:plot-um} but zoomed into the RFI-T1 frequency range. \highlight{The subplot to the right of the waterfall plot illustrates the temporal evolution of RFI-T1 within the 45 – 55 MHz range. The grey line, black dashed line, and shaded light grey region indicate the averaged power spectrum $P^*$, the mean power, and the corresponding 3$\sigma$ confidence levels respectively, all within the said frequency range.}}
    \label{fig:plot-um-rfi-t1}
\end{figure}

\begin{figure*}
    \centering
    \includegraphics[width=0.95\linewidth]{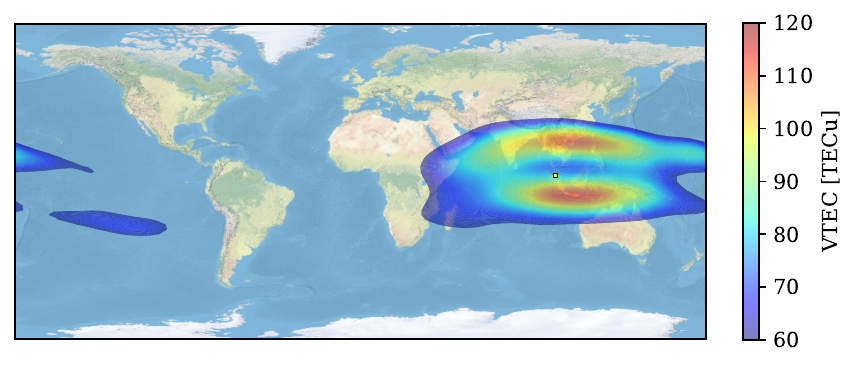}
    \caption{Vertical Total Electron Content (VTEC) map at 3:30 p.m. (GMT+8) on 2 November 2023 from the Chinese Academy of Sciences' real-time global ionospheric map (CAS RT-GIM) model as described in \citealt{Li2020}. The yellow square marks the location of the UM station.}
    \label{fig:tec}
\end{figure*}

RFI-T1 (in the UM data) also consists of an interesting RFI signal centred at 50 MHz with a 10 MHz bandwidth as shown in Figure \ref{fig:plot-um-rfi-t1}, attributed to a newly-used VHF digital communication radio which was previously unseen. Typically, signals within this frequency range traverse through the ionosphere and extend into space, making them unsuitable for skywave propagation of communication signals. However, equatorial ionospheric anomalies (EIA) can augment the reflection of such radio signals \citep{Stening1992, Lin2007, TulasiRam2009, Balan2018}. Interaction between the equatorial daytime electric field and the geomagnetic field drifts the ionospheric plasma upward and away from the dip equator, resulting in a double-peak structure in the plasma density in the north-south direction. In this case, according to a geomagnetic reference frame, the southern hump in the EIA can be seen passing over Malaysia when the emissions are prevalent as shown in Figure \ref{fig:tec}.

Similar to RFI-T1, RFI-T2 (120.12 to 138.18 MHz) exhibits random signals, featuring a higher maximum intensity of -43.23 dBm. This frequency range is designated for aeronautical mobile services or airband, specifically within the \highlight{117.825 to 137 MHz range allocated by the ITU for all regions}, where amplitude modulation voice transmissions (VHF-AM) facilitate communication between aircraft and ground stations, commonly referred to as air traffic control (ATC) services \citep{FAAAC9050D}. \highlight{Under footnote 5.200, ITU also allocates the 121.5 and 123.1 MHz frequencies for aeronautical emergency communications, primarily handling distress and safety calls.}

\section{RFI Mitigation Methods}\label{sec:mitigation}

To continue performing radio observations in environments with high RFI contamination (either due to high population densities such as in Malaysia, and/or due to spectrum allocated to other uses), RFI mitigation techniques have to be implemented. Various mitigation methods can be considered, each with varying complexity and reliability \citep{Fridman2001}. Here we test out two \highlight{post-observation} mitigation methods i.e. removal of RFI corrupted data after \highlight{data collection}, and one physical RFI suppression method, as well as discuss other possible mitigation procedures. 

\subsection{\highlight{Post-Observation} RFI Mitigation}\label{sub:postobservation}

\highlight{Post-observation} RFI mitigation in radio astronomy is highly advantageous due to its versatility, allowing for adjustable flagging thresholds and parameters tailored to the target data quality, all while having minimal to no effect on the instrument's front- and back-end systems. 

\begin{figure}
    \centering
    \includegraphics[width=0.95\linewidth]{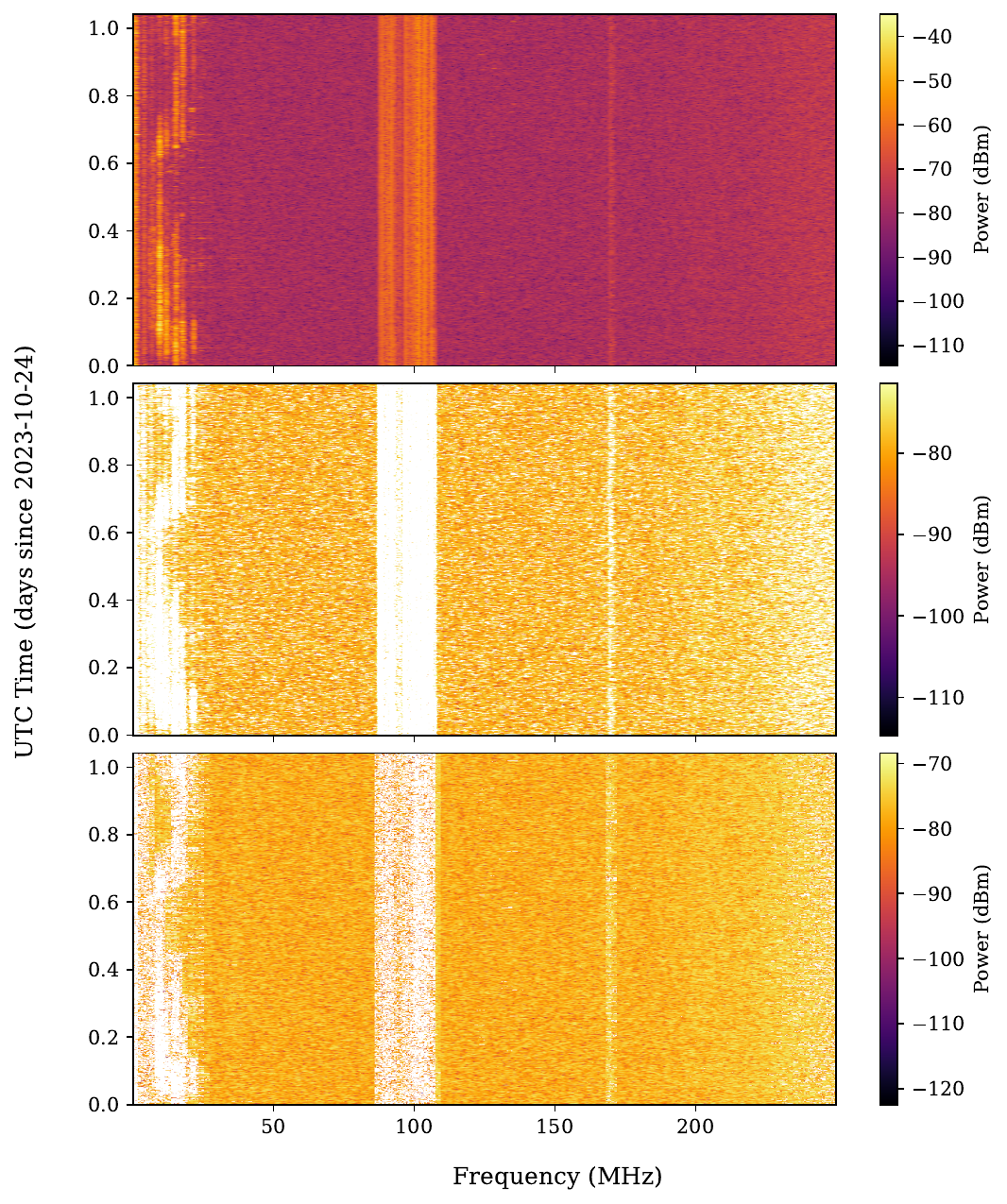}
    \caption{Waterfall plot of the GL data before any RFI mitigation (top panel), after applying the MAD filter (middle panel), and after applying the DWT filter (bottom panel). White pixels denote the time and frequency at which the data has been flagged.}
    \label{fig:plot-mitigation}
\end{figure}

One of the mitigation methods is the Median Absolute Deviation (MAD) filter which utilizes high-order statistics. MAD considers that RFI signals are simply outliers of random Gaussian noise, but utilizes the median deviation instead of the standard deviation which can be influenced by outliers \citep{Buch2016}. MAD can be applied to either the time or frequency domain, but suffers if the RFI contaminates more that 50\% of the data. When applied to UM and GL data in both domains, all RFI were detected well, but in most cases overcompensated (e.g. RFI-P1), including transients. Notably, the gradual positive slope in the signals above 200 MHz was falsely suppressed, posing a threat to data with an uncorrected bandpass shape or a smooth continuum background emission. Shown in Figure \ref{fig:plot-mitigation} and Table \ref{tab:alternative} is the result of the application of the MAD filter on the GL data. 

Another \highlight{post-observation} RFI mitigation method is autocleaning where the RFI waveform is estimated and subtracted from the data. The RFI waveform can be estimated using various filtering techniques including spline-smoothing, wavelet analysis, Wiener filtering, and parametric estimation \citep{Baan2019}. One of which is the discrete wavelet transform (DWT), and this has been shown to be effective in suppressing various types of RFI \citep{Camps2009, DezGarca2019, Tianqi2021}. When a 3-level Haar DWT is applied to both UM and GL data, strong and transient RFI can be seen to be mitigated effectively. Persistent RFI signals are detected, but are mostly undercorrected (e.g. RFI-P2, RFI-P3). Table \ref{tab:alternative} shows the statistics of the DWT performed on the GL data. 

\begin{table}
    \centering
    \caption{Averaged power spectra range in the GL data before any mitigation, and after mitigation using MAD filter and DWT. All values are in dBm.}
    \label{tab:alternative}
    \begin{tabular}{cccc}
        \hline
        Type & Original & MAD filter & DWT \\
        \hline
        Minimum & -114.7737 & -114.7737 & -235.4863 \\
        Maximum & -24.5401 & -71.0455 & -68.0387 \\
        Mean & -75.6161 & -78.4810 & -77.8272 \\
        Std. Dev. & 6.2942 & 3.3162 & 3.9416 \\
        Median & -76.5144 & -77.9297 & -77.3657 \\
        \hline
    \end{tabular}
\end{table}

In recent years, machine learning (ML) has become a powerful tool in astronomy. Techniques such as convolutional neural networks (CNNs), deep fully convolutional networks (DFCNs), and autoencoders have shown the ability to efficiently identify and remove RFI from both simulated and real radio data \citep{Kerrigan2019, Yang2020, DuToit2024}. Supervised ML models trained on labelled data are particularly effective and fast, while unsupervised approaches, which leverage large datasets, offer more robust detection of unknown RFI patterns \citep{Ghanney2020}. ML models can also be trained exclusively on clean, uncontaminated data, allowing them to distinguish RFI from all known astronomical signals and system noise \citep{Mesarcik2022}.

\subsection{Physical RFI Suppression}\label{sub:physicalsuppression}

Physical methods are also viable in suppressing RFI, with the main advantages of real-time application and the usability of the suppressed signals. 

\begin{figure}
    \centering
    \includegraphics[width=0.95\linewidth]{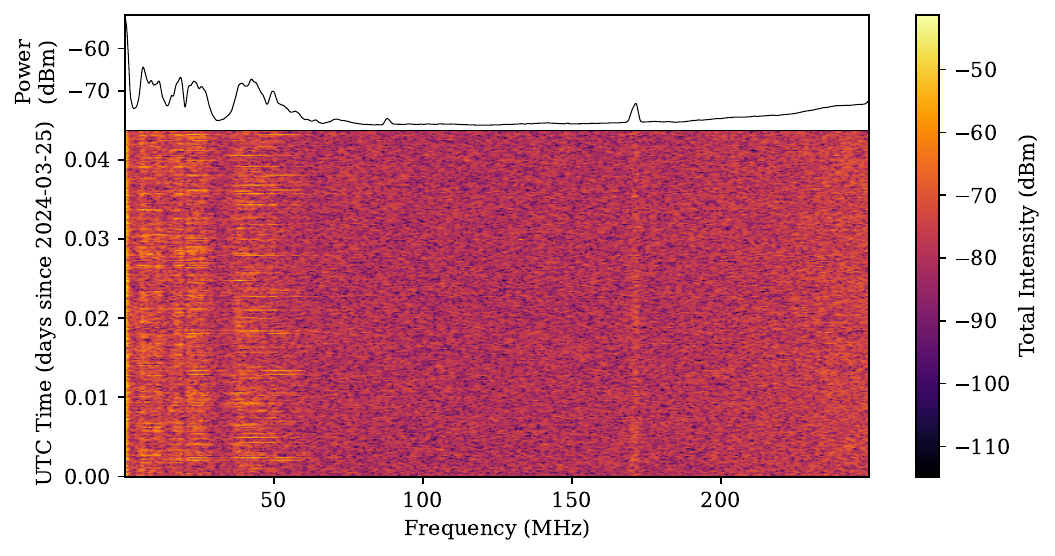}
    \caption{Waterfall plot of a short RFI monitoring test performed at UM using the band stop filter targeted on RFI-P2.}
    \label{fig:plot-filter}
\end{figure}

One of them is using frequency filters which uses priori RFI information to attenuate signals in the contaminated frequency ranges directly from the antenna. The simplest of which is a band stop filter created based on extensive RFI monitoring. Based on our results, we synthesized a band stop filter with a 50 dB attenuation and a roll range between 76 and 122 MHz, and it was able to suppress RFI-P2 in a short RFI monitoring test performed at UM as shown in Figure \ref{fig:plot-filter}. 
\highlight{However, traditional filtering techniques are spectrally inefficient due to their gradual roll-off characteristics, and while superconducting filters offer much sharper selectivity, they are considerably more expensive.} Moreover, the performance of band stop filters depend highly on the quality of the RFI monitoring, and is unable to mitigate transients. To deal with time-varying RFI, adaptive filters which constantly estimate and filter RFI signals within a closed loop \citep{Widrow1985} can be implemented, and has been seen to be effective on various radio telescopes \citep[e.g.][]{Finger2017}.

Another type of physical suppression involves spatial-filtering which performs adaptive suppression of RFI signals based on their direction-of-arrival (DOA), as applied in radar and communication systems \citep{Kocz2010, Steeb2016}. However, this method in its simplest requires an auxiliary reference antenna or multi-feed receivers to capture the RFI DOA, \highlight{which can only be used to overcome single strong interferers and is therefore unusable in complex RFI environments.}

To extend that, cross-correlation data when performing VLBI between telescopes with large baselines is relatively free from local RFI as it emulates the spatial filtering technique. Only signals present in all telescopes -- either from astronomical sources or far-field interference -- would survive the correlation process. Ultimately, combining spatial suppression and \highlight{post-observation} mitigation techniques will be able to give the best outcome in terms of RFI mitigation \citep{Nita2020}.

\highlight{Although our results are based on post-detection processing, it is necessary to acknowledge the benefits of real-time RFI suppression at the intermediate frequency (IF) stage. Strong and narrowband RFI signals can cause spectral leakage, contaminating neighbouring channels even after offline mitigation. Implementing online IF baseband clipping prior to digitalization has been shown to be highly effective in removing such signals at their source \citep[see][]{Fridman2001}. Unfortunately, our instrument setup does not support real-time IF-domain blanking; however, future implementations incorporating FPGA-based backends may offer a promising solution \citep[e.g.][]{QuirsOlozbal2016}.}

\section{Conclusion}\label{sec:conclusion}

We report the measurements of low frequency RFI in two locations in Malaysia; UM representing an urban environment, and GL representing a candidate RNZ. We detected higher levels of RFI in UM than in GL in this frequency range. Using modified GSK, we were able to characterize three groups of persistent RFI and two groups of transient RFI, each attributed to their respective ITU and MCMC spectrum allocations. Some of the RFI detected also exhibit variabilities corresponding to natural phenomena e.g. diurnal frequency evolution due to skywave propagation, and augmentation of skywave propagation by EIA. Shown in Table \ref{tab:summary} is a summary of all the RFI groups detected and their properties.

Various \highlight{post-observation} RFI mitigation and physical RFI suppression techniques were tested on our data as well, and were effective in removing a significant amount of RFI. Ultimately, by combining RNZ characteristics and these mitigation techniques, GL is appropriately capable for relatively low frequency radio astronomy for diffuse radio observations, deep extragalactic surveys, pulsar and FRB monitoring, and solar observations. 

This research can also be viewed as a continuation of the previous RFI monitoring efforts in Malaysia \citep[especially][]{Abidin2021}, step towards the preparation for the development of a research-grade radio telescope in Malaysia. \highlight{It should be noted that the results in this study are based on a single day of observations at each site. While this duration was sufficient to identify key RFI sources and examine short-term patterns such as diurnal variations, it does not capture longer-term or seasonal trends.} Therefore, our future works will therefore include longer monitoring sessions to capture the full range of RFI variability, as well as a larger number of monitoring stations with a sparse spatial distribution to compare a wider range of environmental variability and population differences. 

\begin{table*}
    \centering
    \caption{Summary of low to very high frequency RFI detected using modified GSK in this paper.}
    \label{tab:summary}
    \begin{tabular}{
            p{0.075\linewidth}
            p{0.075\linewidth}
            p{0.125\linewidth}
            p{0.050\linewidth}
            p{0.175\linewidth}
            p{0.325\linewidth}
        }
        \hline
        RFI Type & Identifier & Frequency Range (MHz) & Max. Power (dBm) & Spectrum Allocation & Notes \\
        \hline
        \multirow{4}{*}{Persistent} & RFI-P1 & 0.00 -- 31.25 & -24.54 & Mobile services, broadcasting, maritime, amateur & Constant direct/ground-wave propagation (RFI-P1a) and skywave propagation with diurnal pattern (RFI-P1b) \\
        & RFI-P2 & 86.43 -- 108.40 & -51.63 & Mobile services, broadcasting & FM analogue radio broadcast \\
        & RFI-P3 & 168.21 -- 171.63 & -59.08 & Mobile services & Allocated to Malaysian Government (MLA14) \\
        \hline
        \multirow{4}{*}{Transient} & RFI-T1 & 32.23 -- 56.64 & -51.23 & Mobile services, radiolocation, broadcasting, amateur, space research, radio astronomy & Detection of new VHF signal augmented by EIA skywave propagation \\
        & RFI-T2 & 120.12 -- 138.18 & -43.23 & Aeronautical mobile services & Air traffic control VHF-AM \\
        \hline
    \end{tabular}
\end{table*}

\section*{Acknowledgments}
We acknowledge the support provided by Malaysia’s Ministry of Science, Technology and Innovation (MOSTI) under the Technology Development Fund (TDF07221598). We were also supported by the People's Republic of China's Ministry of Science and Technology (MOST) under the National Key R\&D Programme (SQ2022YFE010485). We finally acknowledge the Chinese Academy of Sciences under the International Partnership Program of the Bureau of International Cooperation of the Chinese Academy of Science and Technology: “Belt and Road” Project (114A11KYSB20200001).

We would like to express our sincere gratitude to the National Astronomical Observatories (NAOC) and Yunnan Astronomical Observatory (YNAO) of the Chinese Academy of Sciences for providing us with the equipments and technical support. We thank the members of the Radio Cosmology Research Laboratory, Universiti Malaya, and the staffs of the Glami Lemi Biotechnology Research Centre, Jelebu, who were involved in the data collection phase of the project.

We greatly appreciate Prof. Gelu M. Nita from the Center for Solar-Terrestrial Research at the New Jersey Institute of Technology for his invaluable guidance on the details of the GSK method. We also thank Ra\'ul D\'iez Garc\'ia from the Universitat Polit\'ecnica de Catalunya for his assistance on the DWT RFI mitigation method. Finally, we gratefully acknowledge the anonymous reviewers for their insightful comments and suggestions, which significantly improved the quality of this article.

\bibliographystyle{jasr-model5-names}
\biboptions{authoryear}
\bibliography{ref}

\end{document}